\documentclass[12pt]{article}

\usepackage{amsmath,amssymb}
\usepackage{graphicx}% Include figure files
\usepackage{dcolumn}% Align table columns on decimal point
\usepackage{bm}% bold math
\usepackage[numbers,super,comma,sort&compress]{natbib}

\begin{document}

% makes references listed with 1., 2., etc.  
  \makeatletter
  \renewcommand\@biblabel[1]{#1.}
  \makeatother

\normalsize

%TITLE HEADER:

{\sffamily
\noindent {\bfseries \Large Equilibrium Statistics of Weakly Slip-Linked
Gaussian Polymer Chains} \\

\noindent {TAKASHI UNEYAMA$^{1,2}$, KAZUSHI HORIO$^{1}$ \\}

%\vspace{-8mm}

\noindent {$^{1}$ Institute for Chemical Research, Kyoto University,
Gokasho, Uji, Kyoto
611-0011, JAPAN \\
$^{2}$ JST-CREST, Gokasho, Uji, Kyoto
611-0011, JAPAN}\\
}

%\vspace{-5mm}

\noindent {\em Dated: \today}\\

\begin{center}
\begin{minipage}{5.5in}

{\bf ABSTRACT:}  
We calculate the free energy and the pressure of a weakly
slip-linked Gaussian polymer chains. We show that the equilibrium
statistics of a slip-linked system is different from one of the
corresponding ideal chain system without any constraints by slip-links.
It is shown that the pressure of a slip-linked system
decreases compared with the ideal system, which implies that slip-linked chains
spontaneously form aggregated cluster like compact structures. These are
qualitatively consistent with previous theoretical analyses or
multi chain simulations.
We also show that repulsive
potentials between chains, which have been phenomenologically utilized
in simulations, can cancel the artificial pressure decrease.


\end{minipage}
\end{center}

%\clearpage

%------------------------------------------------------------------------------
\section*{\sffamily \large INTRODUCTION}

The
motion of polymer chains in melts or solutions are
strongly constrained due to the entanglement
effect\cite{Doi-Edwards-book}, if the degrees of polymerization are large.
Although the entanglement affects dynamic properties drastically, it
does not affect the static properties. For example, in polymer melts,
conformations of polymer chains are well described by the ideal, Gaussian
statistics\cite{deGennes-book} beyond a certain length scale.
Therefore, if we are
interested only on static properties, the Gaussian chain model can be
reasonably utilized.
Actually theories based on the Gaussian statistics achieved great
success to predict static phase behaviours of polymer blends or block
copolymers\cite{Matsen-Bates-1996,Matsen-2002,Kawakatsu-book}.
However, once we consider dynamics of entangled chains, the situation
becomes complicated so much.

It is informative to consider simple toy models which
show non-trivial dynamical behaviours due to topological constraints.
It was reported that molecules composed of infinitely thin rods exhibit
glass transition by increasing the density\cite{Renner-Lowen-Barrat-1995,Obukhov-Kobzev-Perchak-Rubinstein-1997,vanKetel-Das-Frenkel-2005},
while thermodynamically they are ideal gases.
The situation is almost the same for entangled polymers. The
entanglement effects exist even for ideal chains as long as they cannot
cross each other. (Although interactions between monomers or between
monomer and solvent alter rheological behaviours somehow, such effects are
secondary.)
Unfortunately, it is not easy to treat such dynamic effects theoretically.
Thus several phenomenological models such as the tube model\cite{Doi-Edwards-book} or
the slip-link model\cite{Ball-Doi-Edwards-Warner-1981} have been
proposed and utilized.
The slip-link model represent an entanglement as a linking point on
polymers which can slide along the chains (slip-links).
So far, various slip-link type models\cite{Ball-Doi-Edwards-Warner-1981,Schieber-2003,Hua-Schieber-1998,Masubuchi-Takimoto-Koyama-Ianniruberto-Greco-Marrucci-2001,Doi-Takimoto-2003,Nair-Schieber-2006}
have been utilized to study rheological properties of entangled
polymers, and slip-link models have reproduced
various rheological properties well.

However, it is reported that the slip-link models can exhibit
unintuitive
statistics in several situations\cite{Rieger-1987,Rieger-1988,Rieger-1989,Sommer-1992,Metzler-Hanke-Dommersnes-Kantor-Kardar-2002,Metzler-Hanke-Dommersnes-Kantor-Kardar-2002a,Kardar-2008}.
The simplest example is the statistics of a single
Gaussian chain with slip-links.
Rieger\cite{Rieger-1987,Rieger-1988,Rieger-1989} calculated the
partition function of a
Gaussian chain with a slip-link and showed that this system can exhibit
spontaneous symmetry breaking and bifurcation.
Metzler, Kardar and co workers\cite{Metzler-Hanke-Dommersnes-Kantor-Kardar-2002,Metzler-Hanke-Dommersnes-Kantor-Kardar-2002a,Kardar-2008}
studied a single ring polymer with slip-links as a model
of a knotted ring polymer, and
found that slip-linked portion localizes on a ring. This
is similar to Rieger's results.
They reported that the localization behaviour is observed even in
presence of the excluded volume interaction.
Sommer\cite{Sommer-1992} studied two chains bound by a
slip-link, and found similar unintuitive behaviours. He also found that
when multiple chains are bound by multiple slip-links, slip-links can
form compact aggregated structures.
These results imply that various slip-link models may
exhibit unphysical results.

We consider that macroscopic statistical properties of slip-linked
chains may be affected by slip-links. This can be serious problems when
we utilize slip-link models to study various macroscopic properties
(such as pressure or rheology).
As far as the authors know, so
far studies for such unintuitive statistics are limited for simple
systems such as single chain systems.
How slip-links affect macroscopic
thermodynamic properties is not clear.
In principle, if the statistics of the system is explicitly specified
(if the equilibrium probability distribution is explicitly given), we can
check whether the macroscopic thermodynamic properties are really affected or not.
For example, in single chain slip-link models with the mean filed type
approximation\cite{Schieber-2003,Nair-Schieber-2006} or
in virtually coupled many chain slip-link models\cite{Likhtman-2005},
we can straightforwardly show that there are no unintuitive behaviors
(as discussed in Appendix A).

The situation is not simple when we consider multi chain
slip-link models\cite{Masubuchi-Takimoto-Koyama-Ianniruberto-Greco-Marrucci-2001}.
In this work we consider statistics of weakly slip-linked multi
polymer chains theoretically. We calculate the partition function of weakly
slip-linked Gaussian chains by utilizing the grand canonical ensemble
type treatment developed by Schieber\cite{Schieber-2003}.
We show that we can calculate the explicit forms of partition
function and free energy as perturbation expansions.
The free energy of weakly slip-linked chains are shown to be different
from one for ideal chains without any constraints by slip-links.
We also show that slip-links decrease the pressure, and
thus repulsive interactions between chains are required for such systems
to recover (or mimic) the ideal statistics.

\section*{\sffamily \large THEORY}

\subsection*{\sffamily \normalsize Thermodynamic Equilibrium Distribution of Gaussian Chains}

Before we calculate the partition function for slip-linked chains, we
first consider the partition function of Gaussian chains without
any constraints by slip-links. (We call such systems as ``the ideal
chain systems'' or ``the ideal systems'', in the followings.)
As we mentioned, even if chains are well
entangled, equilibrium statistical properties can be well
described by the Gaussian chain model since the entanglement effect is
purely kinetic. As pointed simply by Doi\cite{Doi-book}, the entanglement
effects exist even for ideal chains without excluded volume
effects. Therefore we limit ourselves to ideal chains and ignore
interaction potentials between monomers although they may not be
negligible for some cases (especially for small length scales).
Thus the partition function or the free
energy for ideal chains can be directly applied both to unentangled or entangled
polymers.
Although one may consider this is an oversimplification, as we will show in
the followings, we can still obtain nontrivial results for slip-linked
chains.
We will compare the results of this subsection with the statistics of
slip-linked chains later.

Some expressions in this subsection can be
utilized in the next subsection, to calculate the partition function for
slip-linked chains. Thus here we calculate the partition function for
Gaussian chains rather verbosely.
It is convenient to start from a single chain partition function, since
the partition function of the system can be easily calculated from the
single chain partition function in non-interacting multi chain systems.
We describe the conformation of a polymer chain as $\bm{R}(s)$ ($0 \le s
\le N$). $s$ is
the index along the chain and $N$ is the degree of polymerization.
The single chain
partition function without any constraints can be formally described by
using the Edwards Hamiltonian and the functional integral expression.
\begin{equation}
 \label{partition_function_ideal_free}
 \mathcal{Z}_{1} \equiv \int \mathcal{D}\bm{R} \, \exp
  \left[ - \frac{\mathcal{H}_{0}[\bm{R}(\cdot)]}{k_{B} T} \right]
\end{equation}
where $\mathcal{H}_{0}[\bm{R}(\cdot)]$ is the Edwards Hamiltonian
and $\int \mathcal{D}\bm{R}$ means the functional
integral over $\bm{R}$. We assume that the functional integral $\int \mathcal{D}\bm{R}$
contains an appropriate measure so that the resulting functional integral becomes
dimensionless. The subscript $1$ in the left hand side
of eq \eqref{partition_function_ideal_free} indicates that it is
the partition function for a single chain.
The Edwards Hamiltonian for a Gaussian chain is expressed as
\begin{equation}
 \label{edwards_hamiltonian_ideal}
 \mathcal{H}_{0}[\bm{R}(\cdot)] = \frac{3 k_{B} T}{2 b^{2}} \int_{0}^{N} ds \,
  \left|\frac{\partial \bm{R}(s)}{\partial s }\right|^{2}
\end{equation}
Here $k_{B}$ is the Boltzmann constant, $T$ is the temperature, and $b$
is the segment size. (As we mentioned, the interaction between monomers
are not considered in this work and thus we have no interaction energy
in eq \eqref{edwards_hamiltonian_ideal}.)

We introduce the constraint which fixes the position of the
chain at the index $s_{1}$, $\bm{R}(s_{1})$, to $\bm{r}_{1}$. From
translational symmetry (translational invariance), the partition function under this constraint is
independent of $s_{1}$ and $\bm{r}_{1}$. Then we can define the
following partition function.
\begin{align}
 \label{partition_function_ideal_internal}
 \mathcal{Q} \equiv \int \mathcal{D}\bm{R} \, \left[ \Lambda^{3}
 \delta(\bm{r}_{1} - \bm{R}(s_{1})) \right]
 \exp \left[ - \frac{\mathcal{H}_{0}[\bm{R}(\cdot)]}{k_{B} T} \right]
\end{align}
Here we introduced the dimensional factor $\Lambda^{3}$ (with $\Lambda$
being the thermal de Broglie wave length of the chain) to make the
partition function dimensionless.
Eq \eqref{partition_function_ideal_internal} corresponds to the
partition function for the internal degrees of freedom.
By using eq \eqref{partition_function_ideal_internal}, the single chain
partition function \eqref{partition_function_ideal_free} can be
rewritten as
\begin{equation}
 \label{partition_function_ideal_free_modified}
 \mathcal{Z}_{1} = \frac{1}{N \Lambda^{3}} \int_{0}^{N} ds_{1} \int d\bm{r}_{1} \,
 \mathcal{Q}
 = \frac{V \mathcal{Q}}{\Lambda^{3}}
\end{equation}
where $V$ is the system volume.

Next we consider a canonical ensemble of Gaussian chains.
The partition function for $M$ Gaussian chains can be simply written as
follows, by using the single chain partition function \eqref{partition_function_ideal_free_modified}.
\begin{equation}
 \label{partition_function_ideal}
 \mathcal{Z}^{(0)}
 = \frac{\mathcal{Z}_{1}^{M}}{M!}
 = \frac{V^{M} \mathcal{Q}^{M}}{M! \Lambda^{3 M}}
\end{equation}
The superscript $(0)$ is used to indicate an ideal system.
The free energy of the system is simply calculated to be
\begin{equation}
 \label{free_energy_ideal}
 \mathcal{F}^{(0)} = - k_{B} T \ln \mathcal{Z}^{(0)}
  = k_{B} T M \left[ \ln \frac{M \Lambda^{3}}{V \mathcal{Q}} - 1 \right]
\end{equation}
Thus the pressure becomes as follows.
\begin{equation}
 \label{pressure_ideal}
 P^{(0)} = - \frac{\partial \mathcal{F}^{(0)}}{\partial V}
  = \frac{k_{B} T M}{V}
\end{equation}
This is nothing but the standard van't Hoff form (the equation of state
for ideal gas).
The pressure is simply proportional to the polymer chain density $M / V$.
We again emphasize that eq \eqref{pressure_ideal} holds
even if chains are entangled, because the entanglement effect is purely kinetic.
Although results shown here are rather trivial, it is not trivial
whether the slip-linked chains have the same statistics or not.
In the following sections, we calculate the static thermodynamic properties
of slip-linked chains.

\subsection*{\sffamily \normalsize Partition Function of Slip-Linked Chain Clusters}

\begin{figure}
\begin{center}
\includegraphics[width=0.9\columnwidth,keepaspectratio=true]{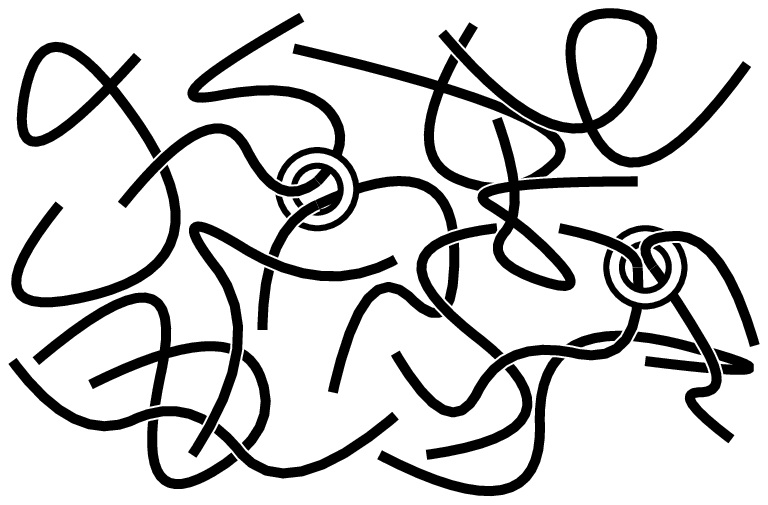}
\end{center}
\caption{\label{weakly_sliplinked_chains_image}
An image of a weakly slip-linked multi chain system. Black curves and
white circles represent polymer chains and slip-links. Only a small
fraction of chains are constrained by slip-links.}
\end{figure}

We consider weakly slip-linked chains in this and following subsections.
For simplicity, we assume that slip-links do not interact each other, and a
single slip-link always bind two chain portions in space.
Under the constraints by slip-links, the thermodynamic properties of the
system will be altered. It can be shown by calculating the partition
function for a slip-linked chains. If we assume that the constraints by
slip-links are sufficiently weak (or chains are sufficiently short),
it is justified to neglect
complicated network like structures with many slip-links.
We show an image of a weakly slip-linked multi chain system in Figure \ref{weakly_sliplinked_chains_image}.
Then we need to calculate only several simple
slip-linked structures, to obtain the expression of the partition
function of the system.
This is similar to the case of weakly interacting real gas\cite{Mayer-Mayer-book}.
If the interaction between gas molecules is sufficiently weak, we only
need several lower order Mayer clusters to study thermodynamic
properties.

The most simple slip-linked structure is two chains paired by one slip-link.
The schematic image of a chain-pair is depicted in Figure
\ref{chain_topologies_entanglement_image}(a).
We may call such a
structure as a ``slip-linked chain cluster'' or simply a ``cluster'', in
analogy to the Mayer clusters.
It is easy to show that the expression for the partition function
is different from the ideal case. The partition function can be
calculated straightforwardly by imposing the slip-linking effect by a
delta function.
We express the conformations of two chains by $\bm{R}_{1}(s)$ and
$\bm{R}_{2}(s)$. Two points on these chains, $\bm{R}_{1}(s_{1})$ and
$\bm{R}_{2}(s_{2})$ are coupled by a slip-link
($s_{1}$ and $s_{2}$ are indices of slip-linked points).
The partition function is calculated as follows.
\begin{equation}
 \label{partition_function_2_1_1}
 \begin{split}
  \mathcal{Z}_{2 (1,1)}
  & = \frac{1}{2! N^{2} \Lambda^{6}} \int \mathcal{D}\bm{R}_{1} \mathcal{D}\bm{R}_{2} \int ds_{1}
  ds_{2} \, [ \Lambda^{3}
  \delta(\bm{R}_{1}(s_{1}) - \bm{R}_{2}(s_{2})) ] \\
  & \qquad \times \exp
  \left[ - \frac{\mathcal{H}_{0}[\bm{R}_{1}(\cdot)] +
  \mathcal{H}_{0}[\bm{R}_{1}(\cdot)]}{k_{B} T}\right] \\
  & = \frac{\mathcal{Q}^{2}}{2! N^{2} \Lambda^{6}} \int_{0}^{N}
  ds_{1} \int_{0}^{N} ds_{2}
  \int d\bm{r}_{1} d\bm{r}_{2} \,
  [\Lambda^{3} \delta(\bm{r}_{1} - \bm{r}_{2})] \\
  & = \frac{V \mathcal{Q}^{2}}{2 \Lambda^{3}}
 \end{split}
\end{equation}
where we introduced the subscript $2 (1,1)$ which indicates that the
cluster is composed of two chains and each chain has one slip-linked
point.

If there are two slip-links, more complicated slip-linked
structures can be formed. In this case, there are three different
possible cluster structures (Figure
\ref{chain_topologies_entanglement_image}(b)-(d)).
The cluster shown in Figure
\ref{chain_topologies_entanglement_image}(b) is composed by three chains.
The clusters shown in \ref{chain_topologies_entanglement_image}(c) and (d)
are composed by two chains, but their topologies are different.
After straightforward calculations, we have the following partition
functions for these clusters.
\begin{equation}
 \label{partition_function_3_1_1_2}
  \mathcal{Z}_{3 (1,1,2)}
   = \frac{V \mathcal{Q}^{3}}{6 \Lambda^{3}}
\end{equation}
\begin{equation}
 \label{partition_function_2_1_3}
 \begin{split}
  \mathcal{Z}_{2 (1,3)}
  & = \frac{\mathcal{Q}^{2}}{2 N^{3} \Lambda^{9}} \int d\bm{r}_{1}
  d\bm{r}_{2} d\bm{r}_{3}
  \int_{0}^{N}
  ds_{1} \int_{s_{1}}^{N} ds_{2} \int_{s_{2}}^{N} ds_{3} \,
   \left(\frac{3 \Lambda^{2}}{2 \pi b^{2}} \right)^{3}
  \frac{1}{[(s_{3} - s_{2}) (s_{2} - s_{1})]^{3/2}} \\
  & \qquad \times \exp \left[ - \frac{3 (\bm{r}_{2} - \bm{r}_{1})^{2}}{2 (s_{2}
  - s_{1}) b^{2}} - \frac{3 (\bm{r}_{3} - \bm{r}_{2})^{2}}{2 (s_{3}
  - s_{2}) b^{2}} \right]
  \Lambda^{3} \delta(\bm{r}_{3} - \bm{r}_{1}) \\
  & = \frac{2 V \mathcal{Q}^{2}}{3 \Lambda^{3}}
  \left(\frac{3 \Lambda^{2}}{2 \pi N b^{2}}\right)^{3/2}
 \end{split}
\end{equation}
\begin{equation}
 \label{partition_function_2_2_2}
 \begin{split}
  \mathcal{Z}_{2 (2,2)}
  & = \frac{\mathcal{Q}^{2}}{2 N^{4} \Lambda^{12}} \int d\bm{r}_{1}
  d\bm{r}_{2} d\bm{r}_{3} d\bm{r}_{4} \int_{0}^{N}
  ds_{1} \int_{s_{1}}^{N} ds_{2} \int_{0}^{N} ds_{3} \int_{s_{3}}^{N} ds_{4} \,
  \left(\frac{3 \Lambda^{2}}{2 \pi b^{2}} \right)^{3} \\
  & \qquad \times \frac{1}{[(s_{4} - s_{3})(s_{2} - s_{1})]^{3/2}}
  \exp \left[ - \frac{3 (\bm{r}_{2} - \bm{r}_{1})^{2}}{2 (s_{2} - s_{1}) b^{2}}
   - \frac{3 (\bm{r}_{4} - \bm{r}_{3})^{2}}{2 (s_{4} - s_{3})
  b^{2}}\right] \\
  & \qquad \times
  \Lambda^{6} \left[ \delta(\bm{r}_{3} - \bm{r}_{1}) \delta(\bm{r}_{4} -
  \bm{r}_{2}) + \delta(\bm{r}_{4} - \bm{r}_{1}) \delta(\bm{r}_{3} -
  \bm{r}_{2}) \right]\\
  & = \frac{16 (7 - 4 \sqrt{2})}{15} \frac{V \mathcal{Q}^{2}}{\Lambda^{3}} \left(\frac{3
  \Lambda^{2}}{2 \pi N b^{2}}\right)^{3/2} \\
 \end{split}
\end{equation}
The subscripts $3 (1,1,2)$, $2 (1,3)$ or $2 (2,2)$ indicate the number of
chains and slip-links on each chain, as before. For example, The cluster $3
(1,1,2)$ is composed by three chains and they have $1$, $1$, and $2$
slip-linked points.

There are other clusters which are composed of
one or two slip-links. As discussed in Appendix B, these clusters have
self-slip-linked structures and such self-slip-linked clusters have no
effect on static thermodynamical properties.
Intuitively, we can understand this as follows.
In the self-slip-linked structures, slip-linked portions
strongly localize\cite{Metzler-Hanke-Dommersnes-Kantor-Kardar-2002,Metzler-Hanke-Dommersnes-Kantor-Kardar-2002a,Kardar-2008}.
Then the partition function for a self-slip-linked structure is almost
the same as the partition function for a structure without
self-slip-links. Thus such clusters practically do not contribute to
the partition function.
Anyway, we do not need to
consider them when we calculate the partition function of slip-linked chains.
All the clusters considered in this subsection are
non self-slip-linked clusters and contribute to thermodynamic properties.

\subsection*{\sffamily \normalsize Equilibrium Free Energy for Weakly Slip-Linked Chains}

We can calculate the partition function of the system from partition
functions for clusters \eqref{partition_function_2_1_1}-\eqref{partition_function_2_2_2}.
We start from a nearly ideal multi chain system where only
small fraction of chains forms slip-linked clusters.
Under such a condition, the system partition function can be calculated
only by using the partition functions for a free chain and a cluster $2 (1,1)$.

According to the Schieber's theory\cite{Schieber-2003}, the
distribution of slip-links can be regarded as a sort of ideal gas in
a grand canonical ensemble, and they are controlled by the effective chemical
potential for slip-links. Then the partition function of the system can be described as
\begin{equation}
 \label{partition_function_multi_chains_definition}
 \mathcal{Z} \equiv \sum_{M_{1} + 2 M_{2} = M}
 \frac{1}{M_{1}! M_{2}!}
  \exp\left( \frac{\epsilon M_{2}}{k_{B} T} \right)
  \mathcal{Z}_{1}^{M_{1}} \mathcal{Z}_{2 (1,1)}^{M_{2}}
\end{equation}
where $M$ is the total number of chains in the system and
$\epsilon$ is the effective chemical potential for a slip-link. The sum in eq
\eqref{partition_function_multi_chains_definition} is taken for $M_{1}$ and
$M_{2}$ under the constraint $M_{1} + 2 M_{2} = M$.
The calculation becomes easier if we consider a grand canonical
ensemble of chains, instead of a canonical ensemble. If we introduce the chemical potential for a chain,
$\mu$, the grand partition function becomes
\begin{equation}
 \label{grand_partition_function_multi_chains_definition}
 \Xi \equiv \sum_{M_{1} = 0}^{\infty} \sum_{M_{2} = 0}^{\infty}
 \frac{1}{M_{1}! M_{2}!}
 \exp\left[ \frac{\mu (M_{1} + 2 M_{2})}{k_{B} T} + \frac{\epsilon M_{2}}{k_{B} T} \right]
  \mathcal{Z}_{1}^{M_{1}} \mathcal{Z}_{2 (1,1)}^{M_{2}}
\end{equation}
Eq \eqref{grand_partition_function_multi_chains_definition} can
be easily calculated.
\begin{equation}
 \label{grand_partition_function_multi_chains_modified}
   \begin{split}
    \Xi
    & = \exp \left[ e^{\mu / k_{B} T} \mathcal{Z}_{1}
    + e^{(2 \mu + \epsilon) / k_{B} T} \mathcal{Z}_{2 (1,1)} \right] \\
    & = \exp \left[ \frac{V \mathcal{Q}}{\Lambda^{3}} e^{\mu / k_{B} T} 
    \left[ 1 + \frac{1}{2} e^{(\mu + \epsilon) / k_{B} T} \mathcal{Q}\right] \right]
   \end{split}
\end{equation}
Thus we obtain the following expression for the grand potential $\mathcal{J}$.
\begin{equation}
 \begin{split}
  \mathcal{J}
  & \equiv - k_{B} T \ln \Xi \\
  & = - k_{B} T  \frac{V \mathcal{Q}}{\Lambda^{3}} e^{\mu / k_{B} T}
    \left[ 1 + \frac{1}{2} e^{(\mu + \epsilon) / k_{B} T} \mathcal{Q}\right]
 \end{split}
\end{equation}
Now we can calculate the free energy by using the Legendre transform.
\begin{equation}
  M = - \frac{\partial \mathcal{J}}{\partial \mu}
  = \frac{V \mathcal{Q}}{\Lambda^{3}}
  e^{\mu / k_{B} T}
    \left[ 1 + e^{(\mu + \epsilon) / k_{B} T} \mathcal{Q}\right]
\end{equation}
\begin{equation}
 \label{free_energy_multi_chains}
 \begin{split}
  \mathcal{F}
  & = \mathcal{J} + \mu M \\
  & = - \frac{1}{2} k_{B} T M
  - \frac{1}{2} k_{B} T M 
  \left[ \frac{1}{2} \sqrt{1 + 4 e^{\epsilon / k_{B} T} \frac{M
  \Lambda^{3}}{V}} + \frac{1}{2} \right]^{-1} \\
  & \qquad - \epsilon M
  + k_{B} T M \ln 
  \left[ \frac{1}{2} \sqrt{1 + 4 e^{\epsilon / k_{B} T} \frac{M \Lambda^{3}}{V}} -
   \frac{1}{2} \right]
  - k_{B} T \ln \mathcal{Q}
 \end{split}
\end{equation}
The pressure of the system is calculated to be
\begin{equation}
 \label{pressure_multi_chains}
  P = - \frac{\partial \mathcal{F}}{\partial V}
  = \frac{k_{B} T M}{2 V}
   \Bigg[ 1 + \left[ \frac{1}{2} \sqrt{1 + 4 e^{\epsilon / k_{B} T}\frac{M \Lambda^{3}}{V}} + \frac{1}{2}  \right]^{-1} \Bigg]
\end{equation}
Clearly the free energy \eqref{free_energy_multi_chains} and the pressure
\eqref{pressure_multi_chains} are different from ones for an ideal
system (eqs \eqref{free_energy_ideal} \eqref{pressure_ideal})

It is convenient to introduce the average number of
slip-linked points on a chain, $\tilde{Z}$.
The average number of slip-linked points
(entanglements) on a chain can be calculated directly
from the free energy.
\begin{equation}
 \label{fraction_of_pairs}
  \tilde{Z}
  = - \frac{2}{M} \frac{\partial \mathcal{F}}{\partial \epsilon}
  = 1 - \left[ \frac{1}{2} \sqrt{1 + 4 e^{\epsilon / k_{B} T}\frac{M \Lambda^{3}}{V}} + \frac{1}{2}  \right]^{-1}
\end{equation}
Here the factor $2$ is introduced because one slip-link corresponds to two
entanglement points (see Figure
\ref{chain_topologies_entanglement_image}).
From eq \eqref{fraction_of_pairs} we find that
the average number of slip-links on a chain depends on the polymer chain
density $M / V$.
We can modify eq \eqref{fraction_of_pairs} and obtain the following
relation.
\begin{equation}
 \label{fraction_of_pairs_modified}
  4 e^{\epsilon / k_{B} T}\frac{M \Lambda^{3}}{V}
  = \bigg(\frac{1 + \tilde{Z}}{1 - \tilde{Z}}\bigg)^{2} - 1
\end{equation}
Eq \eqref{fraction_of_pairs_modified} gives the relation among the
effective chemical potential, the chain number density, and the number of slip-links on a chain.
Since we are considering weakly slip-linked chains, the number of
slip-linked points on a chain should be sufficiently small ($\tilde{Z}
\ll 1$) and thermodynamic quantities
can be expanded into power series of $\tilde{Z}$ around $\tilde{Z} = 0$.
The power series expressions will help us to interpret the physical
meanings of the results.
Eq \eqref{fraction_of_pairs_modified} can be expanded as follows.
\begin{equation}
 \label{fraction_of_pairs_approximated}
  e^{\epsilon / k_{B} T}\frac{M \Lambda^{3}}{V}
  = \tilde{Z} + O(\tilde{Z}^{2})
\end{equation}
By using \eqref{fraction_of_pairs_approximated}, we can rewrite the free energy \eqref{free_energy_multi_chains} and the pressure
\eqref{pressure_multi_chains} simply as follows.
\begin{equation}
 \label{free_energy_multi_chains_approximated}
  \mathcal{F}
  = k_{B} T M \left[ \ln \frac{M \Lambda^{3}}{V \mathcal{Q}}- 1 \right]
  - \frac{1}{2} k_{B} T M \tilde{Z} + O(\tilde{Z}^{2}) \\
\end{equation}
\begin{equation}
 \label{pressure_multi_chains_approximated}
  P = \frac{k_{B} T M}{V} - \frac{1}{2} \frac{k_{B} T M}{V} \tilde{Z}
  + O(\tilde{Z}^{2}) 
\end{equation}
It is now clear that eqs \eqref{free_energy_multi_chains_approximated} and
\eqref{pressure_multi_chains_approximated} reduces to
eqs \eqref{free_energy_ideal} and \eqref{pressure_ideal}, respectively,
at the limit of $\tilde{Z} \to 0$.
We can rewrite eqs \eqref{free_energy_multi_chains_approximated} and
\eqref{pressure_multi_chains_approximated} as follows by using eqs
\eqref{free_energy_ideal} and \eqref{pressure_ideal}.
\begin{equation}
 \label{free_energy_multi_chains_approximated_modified}
  \mathcal{F}
  - \mathcal{F}^{(0)}
  = - \frac{1}{2} k_{B} T M \tilde{Z} + O(\tilde{Z}^{2})
\end{equation}
\begin{equation}
 \label{pressure_multi_chains_approximated_modified}
  P - P^{(0)} = - \frac{1}{2} \frac{k_{B} T M}{V} \tilde{Z} + O(\tilde{Z}^{2}) 
\end{equation}
Eqs \eqref{free_energy_multi_chains_approximated_modified} and
\eqref{pressure_multi_chains_approximated_modified} give the
expressions for the excess free energy and the excess pressure.
From eq \eqref{free_energy_multi_chains_approximated_modified} we find
that the system free energy is decreased by slip-links.
Eq \eqref{pressure_multi_chains_approximated_modified} means that the pressure of
a weakly slip-linked chain system becomes slightly lower than the pressure of the
corresponding ideal chain system.
A weakly slip-linked system does not obey the equation
of state for an ideal gas.

\subsection*{\sffamily \normalsize Effect of Higher Order Clusters}

In the previous subsection, we have considered only the first order cluster.
In this subsection we consider the effect of the higher order clusters.
If we retain the clusters up to the second order, the expression for the
grand potential becomes
\begin{equation}
 \label{grand_partition_function_multi_chains_2nd}
  \begin{split}
   \mathcal{J} & =
   - k_{B} T \bigg[ e^{\mu / k_{B} T} \mathcal{Z}_{1}
   + e^{(2 \mu + \epsilon) / k_{B} T} \mathcal{Z}_{2 (1,1)}
   + e^{(3 \mu + 2 \epsilon) / k_{B} T} \mathcal{Z}_{3 (1,1,1)} \\
   & \qquad + e^{(2 \mu + 2 \epsilon) / k_{B} T} \mathcal{Z}_{2 (1,3)}
   + e^{(2 \mu + 2 \epsilon) / k_{B} T} \mathcal{Z}_{2 (2,2)} \bigg] \\
   & =
   - k_{B} T e^{\mu / k_{B} T} \frac{V \mathcal{Q}}{\Lambda^{3}} \bigg[ 1
   + \frac{1}{2} e^{(\mu + \epsilon) / k_{B} T} \mathcal{Q}
   + \frac{1}{6} e^{(2 \mu + 2 \epsilon) / k_{B} T} \mathcal{Q}^{2} \\
   & \qquad + \frac{2 (61 - 32 \sqrt{2})}{15} e^{(\mu + 2 \epsilon) / k_{B} T}
   \mathcal{Q} \left(\frac{3 \Lambda^{2}}{2 \pi N b^{2}}\right)^{3/2} \bigg]
  \end{split}
\end{equation}
From the results of the previous subsection, we expect that $\xi \equiv
e^{(\epsilon + \mu) / k_{B} T} \mathcal{Q}$ becomes a small perturbation
parameter which represents the strength of
constraints by slip-links. ($\xi$ may be interpreted as the effective
fugacity for slip-links. $\xi = 0$ corresponds to a non slip-linked, free
ideal chain system.)
Then thermodynamic quantities
can be expanded into power series of $\xi$ around $\xi = 0$. Since all the clusters which
contain two or less slip-links are included in eq
\eqref{grand_partition_function_multi_chains_2nd}, we can calculate the
expansion up to $O(\xi^{2})$. From eq
\eqref{fraction_of_pairs_approximated} we find that $O(\tilde{Z}) =
O(\xi)$, and thus we can calculate the expansion series up to $O(\tilde{Z}^{2})$.

The number of chains in the system is calculated to be
\begin{equation}
 \label{chain_number_and_fugacity}
   M =
   e^{\mu / k_{B} T} \frac{V \mathcal{Q}}{\Lambda^{3}} \bigg[ 1
   + \xi 
   + \frac{1}{2} \xi^{2} + \frac{4 (61 - 32 \sqrt{2}) }{15} \xi^{2}
   \frac{e^{- \mu / k_{B} T}}{\mathcal{Q}} \left(\frac{3 \Lambda^{2}
  }{2 \pi N b^{2}}\right)^{3/2} \bigg]
\end{equation}
Then the chemical potential can be related to $M$ and $\xi$ as follows.
\begin{equation}
 \label{chemical_potential_and_fugacity}
  \begin{split}
   \mu
   & = 
   k_{B} T \ln \frac{M \Lambda^{3}}{V \mathcal{Q}}
   - k_{B} T \bigg[ \xi + \frac{4 (61 - 32 \sqrt{2}) }{15} \xi^{2}
   \frac{V}{M} \left(\frac{3}{2 \pi N b^{2}}\right)^{3/2} +
   O(\xi^{3}) \bigg]
  \end{split}
\end{equation}
After the straightforward calculations, we have the following expressions
for the grand partition function or the free energy.
\begin{equation}
   \mathcal{J}
   = 
   - k_{B} T M \bigg[ 1
   - \frac{1}{2} \xi 
   + \frac{2}{3} \xi^{2} - \frac{2 (61 - 32 \sqrt{2}) }{15} \xi^{2}
   \frac{V}{M} \left(\frac{3
  }{2 \pi N b^{2}}\right)^{3/2}
    + O(\xi^{3}) \bigg]
\end{equation}
\begin{equation}
 \label{free_energy_multi_chains_higher_order}
   \mathcal{F}
   = \mathcal{F}^{(0)}
   - k_{B} T M \bigg[
   \frac{1}{2} {\xi}
   + \frac{2}{3} \xi^{2} + \frac{2 (33 - 16 \sqrt{2}) }{15} \xi^{2} \frac{V}{M}
   \left(\frac{3}{2 \pi N b^{2}}\right)^{3/2}  + O(\xi^{3}) \bigg]
\end{equation}
The pressure can be obtained from eq \eqref{free_energy_multi_chains_higher_order}.
\begin{equation}
 \label{pressure_multi_chain_2nd_xi}
   P
   = P^{(0)}
   - k_{B} T \bigg[
   \frac{1}{2} \xi \frac{M}{V}
   + \frac{4}{3} \xi^{2} \frac{M}{V}
   + \frac{2 (61 - 32 \sqrt{2}) }{15} \xi^{2}
   \left(\frac{3}{2 \pi N b^{2}}\right)^{3/2}
   + O(\xi^{3}) \bigg]
\end{equation}
From eq \eqref{pressure_multi_chain_2nd_xi} it is clear that by
increasing $\xi$ (by increasing the strength of the slip-linking
effect), the pressure decreases.

The average number of slip-links $\tilde{Z}$ can be calculated in the same way as
the previous case.
\begin{equation}
 \label{number_of_slip_linked_points_multi_chain}
   \tilde{Z} =
    \xi
   + \frac{8}{3} \xi^{2}
   + \frac{8 (61 - 32 \sqrt{2}) }{15} \xi^{2}
   \frac{V}{M} \left(\frac{3}{2 \pi N b^{2}}\right)^{3/2} + O(\xi^{3})
\end{equation}
Eq \eqref{number_of_slip_linked_points_multi_chain} means that the
number of slip-links monotonically increases as $\xi$ increases, but the
dependence of $\tilde{Z}$ to $\xi$ is nonlinear.
By inverting eq \eqref{number_of_slip_linked_points_multi_chain}, $\xi$
can be expressed as a function of $\tilde{Z}$ as follows.
\begin{equation}
 \label{xi_z_tilde_relation_multi_chain}
   \xi 
   = \tilde{Z}
   - \frac{8}{3} \tilde{Z}^{2} - \frac{8 (61 - 32 \sqrt{2}) }{15} \tilde{Z}^{2}
   \frac{V}{M} \left(\frac{3}{2 \pi N b^{2}}\right)^{3/2} + O(\tilde{Z}^{3})
\end{equation}
Finally the free energy \eqref{free_energy_multi_chains_higher_order}
and the pressure \eqref{pressure_multi_chain_2nd_xi} can be rewritten as
follows, by using the average number of slip-links on a chain, $\tilde{Z}$.
\begin{equation}
 \label{free_energy_multi_chains_higher_order_final}
   \mathcal{F}
   = \mathcal{F}^{(0)}
   - k_{B} T M \bigg[
   \frac{1}{2} \tilde{Z} - \frac{2}{3} \tilde{Z}^{2}  - \frac{2 (61 - 32 \sqrt{2}) }{15} \tilde{Z}^{2}
   \frac{V}{M} \left(\frac{3}{2 \pi N b^{2}}\right)^{3/2} + O(\tilde{Z}^{3}) \bigg]
\end{equation}
\begin{equation}
\label{pressure_multi_chains_2nd}
   P  =
   P^{(0)}
   - k_{B} T \bigg[
   \frac{1}{2} \tilde{Z} \frac{M}{V}
   - \frac{2 (61 - 32 \sqrt{2}) }{15} \tilde{Z}^{2} \left(\frac{3
  }{2 \pi N b^{2}}\right)^{3/2}
   + O(\tilde{Z}^{3})
   \bigg]
\end{equation}
Comparing eqs \eqref{pressure_multi_chains_approximated} and
\eqref{pressure_multi_chains_2nd}, we find that the pressure becomes
slightly lower if we take account of the higher order
clusters. Intuitively this looks inconsistent with eq
\eqref{pressure_multi_chains_2nd} in which the pressure becomes slightly
higher by the higher order clusters.
This comes from the nonlinear
relation between $\xi$ and $\tilde{Z}$ (eq
\eqref{xi_z_tilde_relation_multi_chain}). 
From eqs \eqref{chain_number_and_fugacity} and
\eqref{number_of_slip_linked_points_multi_chain}, $\tilde{Z}$ can be
expanded into power series of $e^{\epsilon / k_{B} T}$ and $M / V$.
\begin{equation}
 \label{number_of_slip_linked_points_multi_chain_final}
 \begin{split}
   \tilde{Z}
  & =
  e^{\epsilon / k_{B} T} \Lambda^{3} \frac{M}{V}
  + \frac{5}{3} e^{2 \epsilon / k_{B} T} \Lambda^{6}
  \left(\frac{M}{V}\right)^{2}
  + \frac{8 (61 - 32 \sqrt{2}) }{15}
  e^{2 \epsilon / k_{B} T} \Lambda^{3}
   \left(\frac{3 \Lambda^{2}}{2 \pi N b^{2}}\right)^{3/2} \frac{M}{V} \\
  & \qquad - \frac{4 (61 - 32 \sqrt{2}) }{15}
  e^{3 \epsilon / k_{B} T} \Lambda^{6}
   \left(\frac{3 \Lambda^{2}
  }{2 \pi N b^{2}}\right)^{3/2} \left(\frac{M}{V}\right)^{2}
  + O(e^{3 \epsilon / k_{B} T} (M / V)^{3})
 \end{split}
\end{equation}
Although eq \eqref{number_of_slip_linked_points_multi_chain_final} is
not simple, we can easily find that $\tilde{Z}$ depends on the chain
density $M / V$ nonlinearly.

\section*{\sffamily \large DISCUSSION}

Here we discuss about the pressure of slip-linked multi chains in detail.
As we have shown, the pressure of a slip-linked system decreases from the
pressure of the corresponding ideal system. Because the thermodynamic control
parameter for slip-linked systems is the chain density $M / V$ and the
effective chemical potential for slip-links $\epsilon$, we rewrite the
pressure \eqref{pressure_multi_chains_2nd} to see how these parameters
affect the pressure.
From eqs \eqref{pressure_multi_chains_2nd} and
\eqref{number_of_slip_linked_points_multi_chain_final}, we can rewrite the excess pressure
as follows.
\begin{equation}
 \label{excess_pressure_multi_chains}
   P - P^{(0)}
   =
   - k_{B} T \bigg[
   \frac{1}{2} e^{\epsilon / k_{B} T} \Lambda^{3}
   + \frac{2 (61 - 32 \sqrt{2}) }{15} e^{2 \epsilon / k_{B} T} \Lambda^{6}
   \left(\frac{3}{2 \pi N b^{2}}\right)^{3/2}
   \bigg] \bigg(\frac{M}{V}\bigg)^{2} + O((M / V)^{3})
\end{equation}
The leading order term of the excess pressure is proportional to $(M / V)^{2}$, and we can interpreter it as the second
order virial.
The second order virial in eq \eqref{excess_pressure_multi_chains} gives
a negative correction to the pressure.
Roughly speaking, the negative second order virial
gives the same contribution as attractive interaction between chains.
Clearly there are no interaction in ideal
systems, and thus the second order virial for an ideal system is
exactly equal to zero. Therefore we can conclude that the negative
second order virial is an artifact of the
slip-link model.
It should be noted that from eq \eqref{excess_pressure_multi_chains} the
excess pressure is proportional to $k_{B} T$. This means that the
origin of the excess pressure is entropic. This can be interpreted that
the entropy loss by slip-links cause the negative virial (or the
effective attraction between chains).

This result implies that we should introduce an
effective repulsive interaction between chains to cancel the artificial
second order virial and recover the correct static thermodynamic properties.
For example, we may introduce a phenomenological repulsive
interaction potential between monomers. If we assume that the interaction
is sufficiently weak, the resulting total pressure, $\tilde{P}$, can be
formally expressed as
\begin{equation}
 \label{effective_pressure_with_monomer_interaction}
 \tilde{P} \approx P + k_{B} T B_{2} \left(\frac{N M}{V}\right)^{2}
\end{equation}
where $B_{2}$ is the second order virial coefficient calculated from the interaction
potential, and $N M / V$ corresponds to the monomer density. By tuning
the interaction potential, we can effectively cancel
the contribution of the artificial second order virial.
The potential should be tuned to satisfy the following
condition. 
\begin{equation}
 \label{second_virial_balance_condition}
 B_{2} \approx
 \frac{1}{N^{2}} \bigg[
   \frac{1}{2} e^{\epsilon / k_{B} T} \Lambda^{3}
   + \frac{2 (61 - 32 \sqrt{2}) }{15} e^{\epsilon / k_{B} T} \Lambda^{6} \left(\frac{3
  }{2 \pi N b^{2}}\right)^{3/2}
   \bigg]
\end{equation}
Notice that, even if the second order virial is successfully cancelled, there are still
higher order virials and the full ideal chain statistics cannot be
reproduced in a slip-linked system.
This situation is somehow
similar to polymer chains in a $\theta$ solvent or real gases at the
Boyle point. Although the second
order virial is cancelled in a $\theta$ solvent, there still
exists a non-zero third or higher order virials.
Therefore generally it is impossible to fully restore the
ideal chain statistics for a slip-linked system, by an artificial interaction
potential.
Similarly, we can introduce an effective interaction between
slip-links. In this case, we have the following total pressure instead of eq \eqref{effective_pressure_with_monomer_interaction}.
\begin{equation}
 \label{effective_pressure_with_sliplink_interaction}
  \tilde{P} \approx P + k_{B} T \tilde{B}_{2} \bigg(\frac{\tilde{Z} M}{V}\bigg)^{2} 
\end{equation}
Here $\tilde{B}_{2}$ is the second order virial for the interaction between slip-links.
By tuning the interaction potential,
the total pressure can be reduced to
the ideal pressure up to the second order in $M / V$, as before.

Here it should be pointed that there is no such an artificial pressure
decrease in the single chain type models, such as slip-linked single
chain models by Schieber and coworkers\cite{Hua-Schieber-1998,Schieber-2003,Nair-Schieber-2006},
the dual slip-link model by Doi and Takimoto\cite{Doi-Takimoto-2003}, or the
slip-spring model by Likhtman\cite{Likhtman-2005}.
In these models, the partition function is essentially the same as the
partition function of ideal chains\cite{Schieber-2003}.
(See Appendix A.)
It is reasonable to consider that
the main reason why we have the artificial
decrease of pressure in a multi chain slip-linked system is that
slip-links bind two chains spatially.
Chains bound by slip-links no longer follow the ideal
statistics and thus the resulting thermodynamic properties become different
from ideal ones.

Let us consider a non-interacting slip-linked multi chain system again,
from a different aspect. Analyses and discussions above are limited for weakly
slip-linked systems.
Here we consider strongly slip-linked systems by utilizing a simple
approximation.
Although in general it is impossible to obtain the analytical expressions
for strongly slip-linked systems, if we consider only limited
sets of slip-linked
clusters (which have simple topologies), it is possible to
obtain approximate expressions.
One of the simplest approximations is to use only
one dimensionally connected, chain-like clusters.
By utilizing this chain clusters approximation, exact expressions for the
free energy or the pressure can be obtained.
Detail calculations are described in Appendix C.
The pressure by the chain cluster approximation becomes as follows.
\begin{equation}
  \label{pressure_chain_cluster_approximation_final}
 P = \bigg( 1 - \frac{\tilde{Z}}{2} \bigg) P^{(0)}
\end{equation}
where $0 \le \tilde{Z} < 2$ is the average number of slip-linked points
on a chain.
Clearly, at the limit of strong slip-linking ($\tilde{Z} \to 2$) the
pressure approaches to zero.
\begin{equation}
 P \to 0 \quad (\tilde{Z} \to 2)
\end{equation}
This can be interpreted that all the chains aggregate into a single
slip-linked cluster, and thus this system is thermodynamically almost
the same as a single ideal gas particle in a large box. The
pressure of such a system is approximately zero, because the pressure is
proportional to the number density of particles and the density is approximately zero.

We expect that the homogeneous state becomes unstable and
slip-linked chains spontaneously form compact aggregates if the
slip-linking effect becomes sufficiently strong.
This is qualitatively in consistent with the theory of Sommer\cite{Sommer-1992}. He predicted that slip-links aggregate even
if there is no other direct interaction between chains.
The compact aggregate formation in slip-linked systems is actually observed in
multi chain slip-link simulations\cite{Masubuchi-Takimoto-Koyama-Ianniruberto-Greco-Marrucci-2001}.
Masubuchi et al\cite{Masubuchi-Takimoto-Koyama-Ianniruberto-Greco-Marrucci-2001}
introduced phenomenological osmotic pressure for their model to
avoid aggregation.
Several different forms of
the phenomenological osmotic pressure have been utilized\cite{Masubuchi-Takimoto-Koyama-Ianniruberto-Greco-Marrucci-2001,Yaoita-Isaki-Masubuchi-Watanabe-Ianniruberto-Greco-Marrucci-2008,Masubuchi-Ianniruberto-Greco-Marrucci-2008,Okuda-Inoue-Masubuchi-Uneyama-Hojo-inpreparation},
and it was found that all of these different forms give qualitatively the same
result. So far, the use of such phenomenological osmotic pressure models are not
theoretically justified.

Our result indicates that osmotic pressure works as effective
repulsion interaction between chains.
If the osmotic pressure is
sufficiently high, the artificial attraction by slip-links can be cancelled.
Detail forms of the repulsive interaction potentials are not essential as long
as they give appropriate positive contributions to the pressure.
This is clear if we consider the condition for a weakly slip-linked
system. There are various
possible potentials which give the same second order virial and satisfy
the condition \eqref{second_virial_balance_condition}.
Therefore we conclude that the detail forms of osmotic pressure are
not important in multi chain slip-link models.
Our analysis justifies the use of
phenomenological repulsive interactions in the multi chain
slip-link simulations.

Here we comment on the stress-optical rule (SOR)\cite{Doi-Edwards-book} in
the multi chain slip-linked
models\cite{Masubuchi-Takimoto-Koyama-Ianniruberto-Greco-Marrucci-2001}.
The rheological properties of multi
chain slip-linked models are usually calculated by using the
SOR. Namely, the stress tensor of the system is assumed to be
determined from the average conformation tensor of subchains between slip-links (or between
a slip-link and a chain end).
In our model, these subchains clearly follow the ideal Gaussian
statistics. The repulsive potential between slip-links do not alter the
chain statistics. Even if we introduce the repulsive potential between monomers,
the interaction between monomers are expected to be screened
if the monomer density is sufficiently large and the interaction range
is sufficiently short\cite{Doi-Edwards-book,deGennes-book,Kawakatsu-book}.
Therefore, the chain statistics is safely assumed to be Gaussian.
If we ignore the contribution from the effective repulsive interaction
between slip-links or monomers, the SOR clearly holds.
However, if we consider the contribution from the repulsive interaction
to the stress tensor, the situation is not simple.
Here we consider the stress tensor of the system, including the
contribution from the repulsive potential between monomers.
The stress tensor of the system can be written as follows.
\begin{equation}
 \label{stress_tensor_multi_chains_monomer_interaction}
\begin{split}
   \bm{\sigma}
 & = \frac{1}{V} \sum_{j} \int_{0}^{N} ds \,
  \frac{3 k_{B} T}{b^{2}} \frac{\partial \bm{R}_{j}(s)}{\partial s}
  \frac{\partial \bm{R}(s)}{\partial s} \\
 & \qquad + \frac{1}{2 V} \sum_{j,j'} \int_{0}^{N} ds ds' \, \frac{\partial
  \tilde{U}_{\text{monomer}}(\bm{R}_{j}(s) - \bm{R}_{j'}(s'))}{\partial
  (\bm{R}_{j}(s) - \bm{R}_{j'}(s'))} (\bm{R}_{j}(s) - \bm{R}_{j'}(s'))
\end{split}
\end{equation}
where $\bm{R}_{j}$ represents the conformation of the $j$-th chain and
$\tilde{U}_{\text{monomer}}(\bm{r})$ is the
repulsive potential between monomers (we assume that
$\tilde{U}_{\text{monomer}}(\bm{r})$ does not diverge at $\bm{r} \to
0$). The second term in the right hand side of eq
\eqref{stress_tensor_multi_chains_monomer_interaction} violates the SOR.
Unfortunately, it is not easy to evaluate how significant this term is,
because it depends on several factors such as the slip-linking strength
or the dynamics model. Recent simulation results show that the
rheological properties (such as the shear relaxation modulus) are not
sensitive to the strength of repulsive 
interaction between monomers, except for the short time scale region\cite{Okuda-Inoue-Masubuchi-Uneyama-Hojo-inpreparation}.
This implies that the contribution from the monomer interaction to the
stress tensor decays rapidly to its equilibrium value.
Then, we expect that eq
\eqref{stress_tensor_multi_chains_monomer_interaction} can reasonably
approximate as the following coarse-grained form, if the slip-linking
effect is not strong and the considered time scale is longer than the
equilibration time of subchains, $\tau_{e}$.
\begin{equation}
 \label{stress_tensor_multi_chains_monomer_interaction_approx}
   \bm{\sigma}
 \approx \frac{1}{V} \sum_{j,k} 
 \frac{3 k_{B} T}{(s_{j,k + 1} - s_{j,k}) b^{2}} (\bm{R}_{j,k + 1} - \bm{R}_{j,k})
  (\bm{R}_{j,k + 1} - \bm{R}_{j,k})
  - k_{B} T B_{2} \left(\frac{N M}{V}\right)^{2} \bm{1}
\end{equation}
where $s_{j,k}$ is the index of the $k$-th slip-link on the $j$-th chain (we
assume $s_{j,1} < s_{j,2} < s_{j,3} < \dots$),
$\bm{R}_{j,k} \equiv
\bm{R}_{j}(s_{j,k})$ is the position of the slip-linked point, and
$\bm{1}$ is the unit tensor.
Therefore, even if we introduce repulsive interactions between
monomers, we can still utilize the SOR (at least as a good
approximation).
If we introduce the effective repulsive potential between slip-links instead of
the potential between monomers, the stress tensor becomes
\begin{equation}
 \label{stress_tensor_multi_chains_slip_link_interaction}
\begin{split}
 \bm{\sigma}
 & = \frac{1}{V} \sum_{j} \int_{0}^{N} ds \,
  \frac{3 k_{B} T}{b^{2}} \frac{\partial \bm{R}_{j}(s)}{\partial s}
  \frac{\partial \bm{R}_{j}(s)}{\partial s} \\
 & \qquad + \frac{1}{2 V} \sum_{j,k \neq j',k'} \frac{\partial
  \tilde{U}_{\text{slip-link}}(\bm{R}_{j,k} - \bm{R}_{j,k})}{\partial
  (\bm{R}_{j,k} - \bm{R}_{j',k'})} (\bm{R}_{j,k} - \bm{R}_{j',k'})
\end{split}
\end{equation}
where $\tilde{U}_{\text{slip-link}}(\bm{r})$ is the repulsive interaction
between slip-linked points.
Again, the second term in the right hand side of eq
\eqref{stress_tensor_multi_chains_slip_link_interaction} violates the
SOR. However, we expect that positions of slip-links will relax
into their local equilibrium positions at the time scale longer than
$\tau_{e}$.
Then for such long time scales eq
\eqref{stress_tensor_multi_chains_slip_link_interaction} will be
approximated as eq
\eqref{stress_tensor_multi_chains_monomer_interaction_approx}, with
$B_{2}$ and $N$ in the second term in the right hand side replaced by
$\tilde{B}_{2}$ and $\tilde{Z}$, respectively, and the SOR becomes
approximately valid.
(Of course such a naive expectation may be incorrect, and if so,
the use of the SOR will not be justified.)
From the discussions above, we consider that the use of the SOR in multi chain slip-link simulations
is practically applicable, except for rather short time regions.
(We do not discuss the rheological properties of our model explicitly,
since the rheological properties depend on the dynamics model and it is
beyond the scope of this work.)

Before we end this section, we shortly comment on cross-linked Gaussian
chain systems\cite{James-Guth-1943,Deam-Edwards-1976,Everaers-1999}.
We expect that our analysis will be also informative
to understand statistical properties of cross-linked chains.
In cross-linked ideal chain systems, chains are connected by
cross-links which do not move along chains.
To calculate the equilibrium statistics, the distribution
function (or statistics) for an index along the chain, $s$, should be
specified. By using the distribution function for an index,
we can calculate the equilibrium statistics in the same way as the slip-linked chains.
We expect that the equilibrium statistics of
randomly cross-linked chains are qualitatively similar to one of
slip-linked chains. Then cross-linked chains are expected to form
compact aggregate like structures.
In fact, it is known that cross-linked chains collapse into compact
structures, without fixing some points on the surface or introducing
periodic boundary conditions\cite{James-Guth-1943,Deam-Edwards-1976,Everaers-1999}.
Our model may be utilized to study several Gaussian chain systems with
various links which bind chains.

\section*{\sffamily \large CONCLUSION}

In this work we calculated the expressions of free energy and
pressure for a weakly slip-linked multi chain system.
Although these results may look physically unnatural, they are consistent
with earlier works which pointed unusual statistics or aggregation
(localization) behaviors in slip-linked systems.
We showed that the free energy or the pressure (the equation of
state) are different from ones for the corresponding ideal
system. This is in contrast to the equilibrium statistics of the single chain
model, which is essentially equivalent to an ideal chain.
Unusual statistics comes from the nature of multi chain slip-link models.

By considering the obtained expression for pressure in detail,
we concluded that slip-links cause the effective interaction between chains.
If the slip-linking effect is sufficiently strong chains spontaneously
aggregate into compact structures.
This is an artifact of the model, and thus we need to introduce repulsive
interactions between monomers or between chains to avoid the unphysical aggregation.
This supports previous reports about multi chain slip-link simulations.

Although our analysis is mainly limited for some simplified cases,
we believe that our results are qualitatively correct even for
strongly slip-linked systems. This work will provide a new strategy to
analyse or improve equilibrium statistics of existing multi chain
slip-link models.

\subsection*{\sffamily \normalsize ACKNOWLEDGMENT}

This work is supported by JST-CREST and the Research Fellowships of the
Japan Society for the Promotion of Science for Young Scientists.
The authors thank Prof. Yuichi Masubuchi for helpful comments.

%------------------------------------------------------------------------------
% appendix

\section*{\sffamily \large APPENDIX A: SINGLE CHAIN MODEL}

In this appendix, we briefly show the results for a weakly slip-linked
single chain model\cite{Schieber-2003}. We consider
equilibrium statistics of slip-linked chains with a sort of mean field
type approximation.
In single chain models, a slip-link does not bind two chains. It is just
placed on a polymer chain to fix it spatially.
As the multi chain model case, we
assume that slip-links are non-interacting.
Then, slip-links are treated as
one dimensional grand canonical ideal gas on a polymer chain. The
effective chemical potential $\bar{\epsilon}$ is used to control the
number of slip-links. (The physical meaning of
$\bar{\epsilon}$ is different from $\epsilon$ in the multi chain
model.)
If the slip-linking effect is sufficiently weak, we can assume that the
number of slip-links on a chain is either $0$ or $1$.
Then the single chain partition
function $\bar{\mathcal{Z}}$ can be written as
\begin{equation}
 \label{single_chain_partition_function_single_chain}
 \bar{\mathcal{Z}} = \frac{V \mathcal{Q}}{\Lambda^{3}}
  (1 + e^{\bar{\epsilon} / k_{B} T})
\end{equation}
The partition function of a many chain system
can be calculated easily by using the
single chain partition function
\eqref{single_chain_partition_function_single_chain}.
In the followings, we also call the ensemble of slip-linked chains with the
mean field type approximation as ``the slip-linked single chain model''
or ``the single chain model'', because its statistics is essentially
determined by the single chain partition function
\eqref{single_chain_partition_function_single_chain}
(although strictly speaking, this expression will not be appropriate).
For the canonical ensemble of chains,
the partition function of the system is expressed as
\begin{equation}
 \mathcal{Z} = \frac{1}{M!} \bar{\mathcal{Z}}^{M}
  = \frac{1}{M!} \left(\frac{V \mathcal{Q}}{\Lambda^{3}}\right)^{M}
  (1 + e^{\bar{\epsilon} / k_{B} T})^{M}
\end{equation}
where $M$ is the number of polymer chains in the system.
Thus the free energy and pressure can be expressed as follows
\begin{equation}
  \mathcal{F}
  = k_{B} T M
  \left[ \ln \frac{M \Lambda^{3}}{V \mathcal{Q}}
   - \ln (1 + e^{\bar{\epsilon} / k_{B} T}) - 1 \right]
\end{equation}
\begin{equation}
 \label{pressure_single_chain}
 P = \frac{k_{B} T M}{V}
\end{equation}
The pressure of the single chain model is thus independent of the
effective chemical potential. From eq \eqref{pressure_ideal}, we find
that the pressure of a slip-linked single chain model is the same as
the pressure of an ideal system.

The average number of slip-links on a chain, $\tilde{Z}$, is given by
\begin{equation}
 \label{number_of_slip_linked_points_single_chain}
 \tilde{Z} = - \frac{1}{M} \frac{\partial \mathcal{F}}{\partial \bar{\epsilon}}
  = \frac{1}{1 + e^{- \bar{\epsilon} / k_{B} T}}
\end{equation}
By inverting eq \eqref{number_of_slip_linked_points_single_chain}, the
effective chemical potential $\bar{\epsilon}$ can be related to
$\tilde{Z}$ as
\begin{equation}
 e^{\bar{\epsilon} / k_{B} T} = \frac{\tilde{Z}}{1 - \tilde{Z}} = \tilde{Z} + O(\tilde{Z}^{2})
\end{equation}
where we used $\tilde{Z} \ll 1$. Finally we have the following expression for
the free energy.
\begin{equation}
 \label{free_energy_single_chain_approximated}
  \mathcal{F}
  = \mathcal{F}^{(0)}
  - k_{B} T M \tilde{Z} + O(\tilde{Z}^{2})
\end{equation}
This expression for the free energy
\eqref{free_energy_single_chain_approximated} looks the same as the free
energy of the multi chain model
\eqref{free_energy_multi_chains_approximated} except for the
numerical coefficient ($1 / 2$). However, in the single chain model,
the average number of slip-links on a chain, $\tilde{Z}$, is independent of the
chain density $M / V$. Therefore the pressure
\eqref{pressure_single_chain} is exactly the same as the pressure of
ideal chains. This means that in the single chain model, the static
thermodynamic properties are the same as the corresponding ideal system.
It should be noted that the pressure is never affected by slip-links even
if we consider strongly slip-linked cases ($\tilde{Z} \gg 1$).

Intuitively, the differences between the single and multi chain models
come from the difference of the slip-linking effects.
Slip-links in the single chain model do not bind
chains while slip-links in the multi chain model do.
This nature is independent of the slip-linking strength, and thus
the discussions in this appendix also holds qualitatively for strongly
slip-linked systems.

We should note that even in multi chain systems, the situation becomes almost
the same if there are no direct spatial coupling between two slip-linked chains.
For example, in
the dual slip-link model by Doi and Takimoto\cite{Doi-Takimoto-2003} or
in the slip-spring model by Likhtman\cite{Likhtman-2005}, slip-linked two
points are coupled ``virtually'' and not directly coupled in space.
The statistics of a chain is not affected by slip-links.
Therefore in these models slip-links do not alter equilibrium statistics
of individual chains, and the resulting expressions of free energy
or pressure (or other static thermodynamical properties) are essentially the same as eqs
\eqref{free_energy_single_chain_approximated} and
\eqref{pressure_single_chain}.
In a sense, we can interpret these virtual coupling type models as
single chain models.

\section*{\sffamily \large APPENDIX B: EFFECT OF SELF-SLIP-LINKS}

We can construct clusters which contain self-slip-links (slip-links
which constraint the same chain).
As mentioned in the main text, the self-slip-linked clusters do not
contribute to equilibrium statistics.
In this appendix, we show that self-slip-linked
can be neglected safely. We calculate the statistical probabilities for
self-slip-linked and show that self-slip-linked spontaneously shrink
and finally disappear in equilibrium.

\subsection*{\sffamily \normalsize First Order Self-Slip-Linked Cluster}

We start from the most simple self-slip-linked structure.
We consider the first order self-slip-link, which is a single
isolated loop like structure (Figure
\ref{chain_topologies_self_entanglement_image}(a)).
We express the indices of slip-linked points by $s_{1}$ and $s_{2}$, and
assume $s_{1} \le s_{2}$.
The partition function can be formally expressed as
\begin{equation}
 \label{partition_function_self_entanglement_1}
 \begin{split}
 \mathcal{Z}_{1 (2)}
  & = \frac{\mathcal{Q}}{\Lambda^{6} N^{2}}
  \int d\bm{r}_{1} d\bm{r}_{2} \int_{0}^{N} ds_{1} \int_{s_{1}}^{N} ds_{2} \,
  \left[\frac{3 \Lambda^{2}}{2 \pi (s_{2} - s_{1}) b^{2}} \right]^{3/2}
  \exp \left[ - \frac{3  (\bm{r}_{1} - \bm{r}_{2})^{2}}{2 (s_{2} - s_{1}) b^{2}}\right]
  \Lambda^{3} \delta(\bm{r}_{2} - \bm{r}_{1}) \\
  & = \frac{V \mathcal{Q}}{\Lambda^{3} N^{2}} \int_{0}^{N} ds_{1}
  \int_{s_{1}}^{N} ds_{2}
  \, \left[\frac{3 \Lambda^{2}}{2 \pi (s_{2} - s_{1}) b^{2}} \right]^{3/2}
 \end{split}
\end{equation}
(As before, the subscript $1 (2)$ in eq
\eqref{partition_function_self_entanglement_1} indicates that the
cluster is composed of one chain and two points are slip-linked.)
However, the integral in eq \eqref{partition_function_self_entanglement_1}
does not converge, due to the divergence of the integrand
at $s_{1} = s_{2}$. Then we may interpret that
the expression \eqref{partition_function_self_entanglement_1} itself has
no physical meaning. Instead of the partition function
\eqref{partition_function_self_entanglement_1}, we consider the
probability for given $s_{1}$ and $s_{2}$,
$\mathcal{P}_{1 (2)}(s_{1},s_{2})$.
This probability is proportional to the
integrand of the last integral eq \eqref{partition_function_self_entanglement_1}.
\begin{equation}
 \label{probability_distribution_self_entanglement_1}
  \begin{split}
 \mathcal{P}_{1 (2)}(s_{1},s_{2})
   & \propto \int d\bm{r}_{1} d\bm{r}_{2} \,
  \left[\frac{3 \Lambda^{2}}{2 \pi (s_{2} - s_{1}) b^{2}} \right]^{3/2}
  \exp \left[ - \frac{3  (\bm{r}_{1} - \bm{r}_{2})^{2}}{2 (s_{2} -
   s_{1}) b^{2}}\right]
  \Lambda^{3} \delta(\bm{r}_{2} - \bm{r}_{1}) \\
   & \propto \frac{1}{(s_{2} - s_{1})^{3/2}}   
  \end{split}
\end{equation}
From eq \eqref{probability_distribution_self_entanglement_1} we obtain
the following effective free energy for $s_{1}$ and $s_{2}$.
\begin{equation}
 \label{free_energy_self_entanglement_1}
 \mathcal{F}_{1 (2)}(s_{1},s_{2})
 = - k_{B} T \ln \mathcal{P}_{1 (2)}(s_{1},s_{2})
  = \frac{3}{2} k_{B} T \ln (s_{2} - s_{1}) + \text{(const.)}
\end{equation}
The effective free energy \eqref{free_energy_self_entanglement_1}
diverges to negative infinity at $s_{1} = s_{2}$, and thus we know that this
system has no lower bound for the free energy.
The system will be trapped at the state $s_{1} = s_{2}$ and
cannot escape from that state. 
In other words, this system has no
thermodynamic equilibrium state and therefore the partition function
\eqref{partition_function_self_entanglement_1} has no physical meaning.
Since the state $s_{1} = s_{2}$
corresponds to unentangled chain, we conclude that the self-slip-link is
spontaneously eliminated.
(These results are essentially the same
as the theories by Sommer\cite{Sommer-1992}, or by Metzler
et al\cite{Metzler-Hanke-Dommersnes-Kantor-Kardar-2002,Metzler-Hanke-Dommersnes-Kantor-Kardar-2002a,Kardar-2008}.)

\subsection*{\sffamily \normalsize Second Order Self-Slip-Linked Clusters with One Chain}

\begin{figure}
\begin{center}
\includegraphics[width=0.9\columnwidth,keepaspectratio=true]{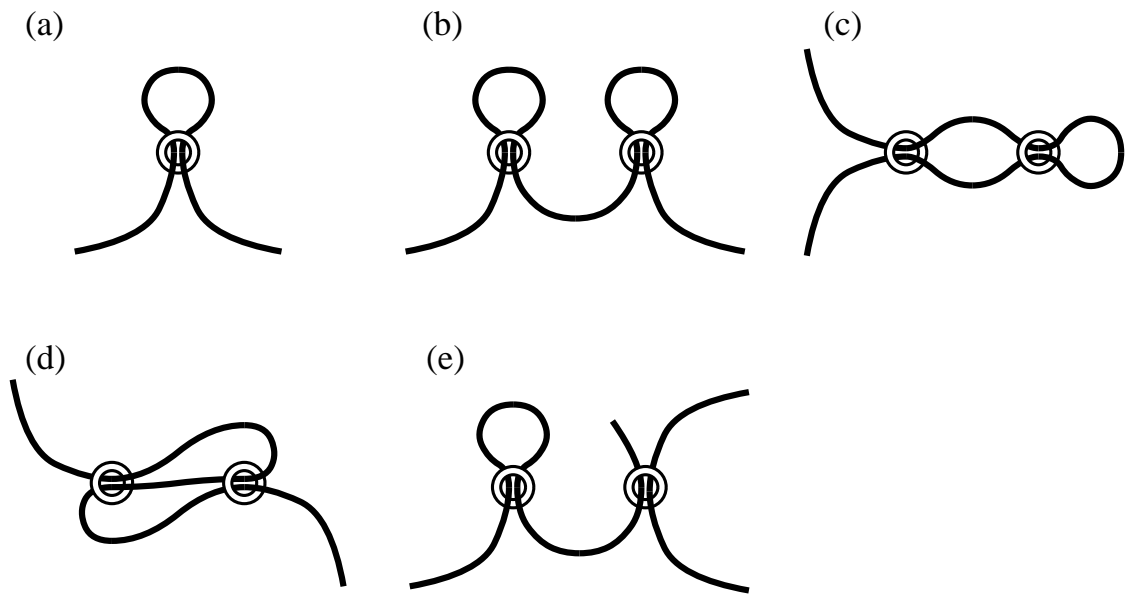}
\end{center}
\caption{\label{chain_topologies_self_entanglement_image}
Schematic images for self-slip-linked polymer chains.
 Black curves and white circles represent polymer chains and slip-links, respectively.
 (a) a chain with an isolated loop ($1 (2)$), (b) a chain with two
 isolated loops ($1 (4)$),
 (c) a chain with nested (``8''-shaped) loops ($1 (4')$), (d) a chain with fused
 (``$\theta$''-shaped) loops ($1 (4'')$), and (e) paired two chains with
 an isolated loop ($2 (3,1)$).}
\end{figure}

We proceed to more complicated cases. Here we consider the second order
self-slip-linked structures which are composed only by one chain.
There are three different
second order self-slip-linked structures; (1) two isolated loops (Figure
\ref{chain_topologies_self_entanglement_image}(b)), (2) nested
loops which has a shape of ``8'' (Figure
\ref{chain_topologies_self_entanglement_image}(c)), and (3) fused loops
which has a shape of ``$\theta$'' (Figure
\ref{chain_topologies_self_entanglement_image}(d)).
We may index clusters shown in Figure
\ref{chain_topologies_self_entanglement_image}(b), (c), and (d) as $1
(4)$, $1 (4')$, and $1 (4'')$, respectively (primes are introduced to
distinguish clusters which have the same number of slip-linked points).

For the case of two isolated loops, the situation is almost the same as
the case of an isolated loop. There are four slip-linked points. We
write indices as $s_{1}, s_{2}, s_{3}$, and $s_{4}$. As before, we
assume $0 \le s_{1} \le s_{2} \le s_{3} \le s_{4} \le N$.
The probability distribution for
indices of slip-linked points can be expressed as
\begin{equation}
 \label{probability_distribution_self_entanglement_2_1}
  \begin{split}
   \mathcal{P}_{1 (4)}(s_{1},s_{2},s_{3},s_{4})
   & \propto \int d\bm{r}_{1} d\bm{r}_{2} d\bm{r}_{3} d\bm{r}_{4} \,
   \frac{1}{[(s_{2} - s_{1}) (s_{3} - s_{2}) (s_{4} - s_{3})]^{3/2}} \\
   & \qquad \times \exp \left[ - \frac{3 (\bm{r}_{2} - \bm{r}_{1})^{2}}{2 (s_{2} -
   s_{1}) b^{2}} - \frac{3 (\bm{r}_{3} - \bm{r}_{2})^{2}}{2 (s_{3} -
   s_{2}) b^{2}} - \frac{3 (\bm{r}_{4} - \bm{r}_{3})^{2}}{2 (s_{4} -
   s_{3}) b^{2}} \right] \\
   & \qquad \times \Lambda^{6} \delta(\bm{r}_{2} - \bm{r}_{1}) 
   \delta(\bm{r}_{4} - \bm{r}_{3}) \\
   & \propto \frac{1}{[(s_{2} - s_{1}) (s_{4} -
   s_{3})]^{3/2}}
  \end{split}
\end{equation}
and the effective free energy becomes
\begin{equation}
 \label{free_energy_self_entanglement_2_1}
 \mathcal{F}_{1 (4)}(s_{1},s_{2},s_{3},s_{4})
  = \frac{3}{2} k_{B} T
  \left[ \ln (s_{2} - s_{1}) + \ln (s_{4} - s_{3}) \right] + \text{(const.)}
\end{equation}
The effective free energy diverges at $s_{2} = s_{1}$ and/or $s_{4} =
s_{3}$. This means that the system will be trapped at the state $s_{2} =
s_{1}$ and $s_{4} = s_{3}$, and thus the loops are spontaneously eliminated.
The current argument can be generalized easily to many isolated loops,
and thus all isolated loops are spontaneously eliminated.

For the case of ``8''-shaped loops, the situation is also similar.
We can calculate the probability distribution of $\lbrace s_{i} \rbrace$
as
\begin{equation}
 \label{probability_distribution_self_entanglement_2_2}
  \begin{split}
   \mathcal{P}_{1 (4')}(s_{1},s_{2},s_{3},s_{4})
   & \propto \int d\bm{r}_{1} d\bm{r}_{2} d\bm{r}_{3} d\bm{r}_{4} \,
   \frac{1}{[(s_{2} - s_{1}) (s_{3} - s_{2}) (s_{4} - s_{3})]^{3/2}} \\
   & \qquad \times \exp \left[ - \frac{3 (\bm{r}_{2} - \bm{r}_{1})^{2}}{2 (s_{2} -
   s_{1}) b^{2}} - \frac{3 (\bm{r}_{3} - \bm{r}_{2})^{2}}{2 (s_{3} -
   s_{2}) b^{2}} - \frac{3 (\bm{r}_{4} - \bm{r}_{3})^{2}}{2 (s_{4} -
   s_{3}) b^{2}} \right] \\
   & \qquad \times \Lambda^{6} \delta(\bm{r}_{3} - \bm{r}_{2})
   \delta(\bm{r}_{4} - \bm{r}_{1})\\
   & \propto \frac{1}{[(s_{3} - s_{2}) (s_{4} - s_{3} + s_{2}
   - s_{1})]^{3/2}}
  \end{split}
\end{equation}
The effective free energy is expressed as
\begin{equation}
 \label{free_energy_self_entanglement_2_2}
 \mathcal{F}_{1 (4')}(s_{1},s_{2},s_{3},s_{4})
  = \frac{3}{2} k_{B} T
  \left[ \ln (s_{3} - s_{2}) + \ln (s_{4} - s_{3} + s_{2} - s_{1}) \right] + \text{(const.)}
\end{equation}
As before, the effective free energy diverges at $s_{3} = s_{2}$ and/or
$s_{4} = s_{3} - s_{2} + s_{1}$. Thus the system will be trapped at the
state $s_{3} = s_{2}$ and $s_{4} = s_{1}$. Thus we find that the
``8''-shaped loops are also spontaneously eliminated.

Finally we consider the ``$\theta$''-shaped loops case. Although the method
itself is the same as previous cases, the calculation becomes a bit
complicated for this case. We have the following
probability distribution for indices.
\begin{equation}
 \label{probability_distribution_self_entanglement_2_3}
  \begin{split}
   \mathcal{P}_{1 (4'')}(s_{1},s_{2},s_{3},s_{4})
   & \propto \int d\bm{r}_{1} d\bm{r}_{2} d\bm{r}_{3} d\bm{r}_{4} \,
   \frac{1}{[(s_{2} - s_{1}) (s_{3} - s_{2}) (s_{4} - s_{3})]^{3/2}} \\
   & \qquad \times \exp \left[ - \frac{3 (\bm{r}_{2} - \bm{r}_{1})^{2}}{2 (s_{2} -
   s_{1}) b^{2}} - \frac{3 (\bm{r}_{3} - \bm{r}_{2})^{2}}{2 (s_{3} -
   s_{2}) b^{2}} - \frac{3 (\bm{r}_{4} - \bm{r}_{3})^{2}}{2 (s_{4} -
   s_{3}) b^{2}} \right] \\
   & \qquad \times \Lambda^{6} \delta(\bm{r}_{3} - \bm{r}_{1}) 
   \delta(\bm{r}_{4} - \bm{r}_{2}) \\
   & \propto \frac{1}{[(s_{2} - s_{1}) (s_{3} - s_{2}) + (s_{2} - s_{1})
   (s_{4} -
   s_{3}) + (s_{3} - s_{2})(s_{4} - s_{3})]^{3/2}}
  \end{split} 
\end{equation}
and the effective free energy becomes
\begin{equation}
 \label{free_energy_self_entanglement_2_3}
  \begin{split}
 \mathcal{F}_{1 (4'')}(s_{1},s_{2},s_{3},s_{4})
   & = \frac{3}{2} k_{B} T
  \ln \left[ (s_{2} - s_{1}) (s_{3} - s_{2}) + (s_{2} - s_{1})
   (s_{4} - s_{3}) + (s_{3} - s_{2})(s_{4} - s_{3}) \right] \\
   & \qquad + \text{(const.)}
  \end{split}
\end{equation}
The free energy diverges at
\begin{equation}
 \label{trapped_state_condition_self_entanglement_2_3}
 (s_{2} - s_{1}) (s_{3} - s_{2}) + (s_{2} - s_{1})
   (s_{4} - s_{3}) + (s_{3} - s_{2})(s_{4} - s_{3}) = 0
\end{equation}
However, unlike the previous cases, it is not clear whether the
condition \eqref{trapped_state_condition_self_entanglement_2_3}
corresponds to the unentangled state or not.
To reduce the degrees of freedom, we try to integrate eq
\eqref{probability_distribution_self_entanglement_2_3} over $s_{2}$.
\begin{equation}
 \label{probability_distribution_self_entanglement_2_3_modified}
  \begin{split}
   \mathcal{P}_{1 (4'')}(s_{1},s_{3},s_{4})
   & = \int_{s_{1}}^{s_{3}} ds_{2} \, \mathcal{P}_{1 (4'')}(s_{1},s_{2},s_{3},s_{4}) \\
   & \propto \frac{1}{[4 (s_{4} - s_{3}) + (s_{3} - s_{1})]\sqrt{(s_{4} - s_{3})(s_{3} - s_{1})}} \\
  \end{split} 
\end{equation}
The corresponding
effective free energy is
\begin{equation}
 \label{free_energy_self_entanglement_2_3_modified}
  \begin{split}
 \mathcal{F}_{1 (4'')}(s_{1},s_{3},s_{4})
   & = \frac{1}{2} k_{B} T
  \left[ 2 \ln [4 (s_{4} - s_{3}) + (s_{3} - s_{1})]
   + \ln (s_{4} - s_{3})
   + \ln (s_{3} - s_{1}) \right] \\
   & \qquad + \text{(const.)}
  \end{split}
\end{equation}
This free energy diverges at $s_{3} = s_{1}$ and/or $s_{4} =
s_{1}$. Thus we find that the system will be trapped at $s_{4} = s_{3} =
s_{1}$. Similarly, we have the following expression for the effective free energy by
integrating eq \eqref{probability_distribution_self_entanglement_2_3} over $s_{3}$.
\begin{equation}
 \label{free_energy_self_entanglement_2_3_modified2}
  \begin{split}
 \mathcal{F}_{1 (4'')}(s_{1},s_{2},s_{4})
   & = \frac{1}{2} k_{B} T
  \left[ 2 \ln [4 (s_{2} - s_{1}) + (s_{4} - s_{2})]
   + \ln (s_{2} - s_{1})
   + \ln (s_{4} - s_{2}) \right] \\
   & \qquad + \text{(const.)}
  \end{split}
\end{equation}
Eq \eqref{free_energy_self_entanglement_2_3_modified2} gives $s_{4} =
s_{2} = s_{1}$ as the trapped
state. Therefore we find that for the case of
``$\theta$''-shaped loops, the system is trapped at the state $s_{4} =
s_{3} = s_{2} = s_{1}$, and the loops are spontaneously eliminated just
like other cases.

\subsection*{\sffamily \normalsize Second Order Self-Slip-Linked Cluster with Two Chains}

There is another second order self-slip-linked structure, which is
composed by two chains (Figure
\ref{chain_topologies_self_entanglement_image}(e)).
This cluster has the similar structure with clusters $1(2)$ or $1(4)$,
and the calculation is straightforward.

We express the indices for slip-linked points on one chain as $s_{1},
s_{2}$, and $s_{3}$ ($s_{1} \le s_{2} \le s_{3}$), and the index on
another chain as $s_{4}$.
Integrating over all variables except $s_{2}$ and $s_{1}$, we have the
following probability distribution function.
\begin{equation}
 \label{probability_distribution_self_entanglement_2_3_1}
  \begin{split}
   \mathcal{P}_{2 (3,1)}(s_{1},s_{2})
   & \propto \frac{N - s_{2}}{(s_{2} - s_{1})^{3/2}}
  \end{split}
\end{equation}
The corresponding free energy is
\begin{equation}
 \label{free_energy_self_entanglement_2_3_1}
 \mathcal{F}_{2 (3,1)}(s_{1},s_{2})
   = \frac{1}{2} k_{B} T
   \left[ 3 \ln (s_{2} - s_{1})
   - \ln (N - s_{2}) \right] + \text{(const.)}
\end{equation}
As before, the system is trapped at $s_{1} = s_{2}$. Thus the cluster
$2 (3,1)$ reduces to the cluster $2 (1,1)$ (Figure
\ref{chain_topologies_entanglement_image}(a)), which is already
calculated in the main text.

From calculations in this appendix, all the self-slip-linked
clusters up to the second order reduce to non self-slip-linked clusters.
It is natural to expect that all the higher order self-slip-linked
clusters can also be reduced to non self-slip-linked clusters.
Therefore we do not consider the contributions from the
self-slip-linked clusters.

\section*{\sffamily \large APPENDIX C: CHAIN CLUSTERS APPROXIMATION}

\begin{figure}
\begin{center}
\includegraphics[width=0.9\columnwidth,keepaspectratio=true]{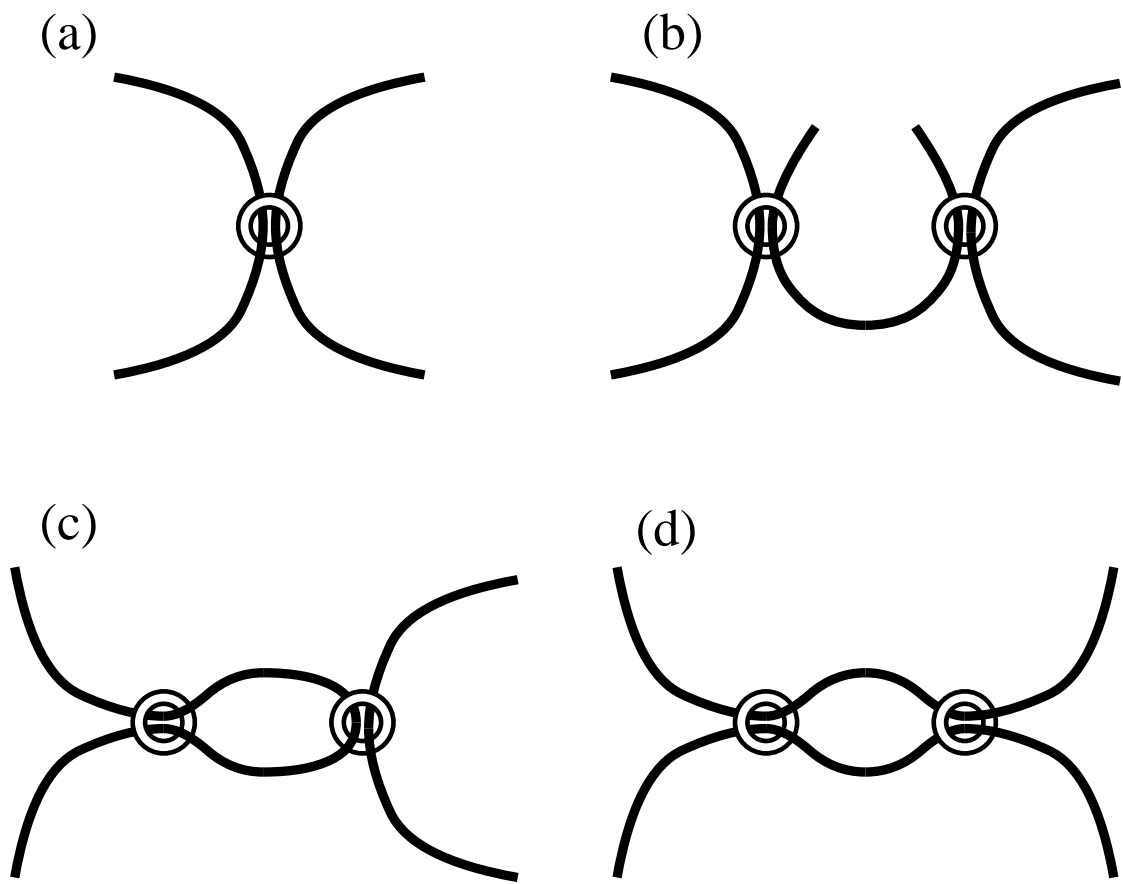}
\end{center}
\caption{\label{chain_topologies_entanglement_image}
Schematic images for polymer chains paired by slip-links.
Black curves and white circles represent polymer chains and slip-links, respectively.
(a) two chains paired by a slip-link ($2 (1,1)$),
(b) three chains coupled by two slip-links ($3 (1,1,2)$), and
(c) and (d) two chains paired by two slip-links ($2 (1,3)$ and $2 (2,2)$).}
\end{figure}

In this appendix we derive expressions for free energy or pressure
with arbitrary slip-linking strength, by utilizing an approximation.
To make the partition function tractable, we limit ourselves to consider
only limited set of clusters, which have one dimensionally
connected, chain-like slip-linked structures (there are no branches or
loops). We may call such clusters as chain
clusters. The $n$-th order chain cluster has $n$ chains and $n - 1$
slip-links. The first order chain cluster is a free chain, and
the second and third chain clusters correspond to $2(1,1)$ (Figure
\ref{chain_topologies_entanglement_image}(a)) and
$3(1,1,2)$ (Figure \ref{chain_topologies_entanglement_image}(b)),
respectively.
We can calculate the partition function of the $n$-th order chain
cluster $n(1,1,2,\dots,2)$ as
\begin{equation}
 \mathcal{Z}_{n (1,1,2,\dots,2)} = \frac{V \mathcal{Q}^{n}}{n! \Lambda^{3}}
\end{equation}
It is then straightforward to calculate the grand partition function or
the grand potential for chain clusters.
\begin{equation}
 \mathcal{J}
  = - k_{B} T \sum_{n = 1}^{\infty}
   e^{[n \mu + (n - 1) \epsilon] / k_{B} T} \mathcal{Z}_{n (1,1,2,\dots,2)}
  = - k_{B} T e^{- \epsilon / k_{B} T} \frac{V}{\Lambda^{3}} \left[ e^{\xi} - 1 \right]
\end{equation}
where we defined $\xi \equiv e^{(\mu + \epsilon) / k_{B} T}
\mathcal{Q}$.
The number of chains $M$ is calculated to be
\begin{equation}
 \label{number_of_chains_chain_cluster_approximation}
 M
  = e^{- \epsilon / k_{B} T} \frac{V}{\Lambda^{3}} \xi e^{\xi}
\end{equation}
Thus the free energy can be expressed as follows.
\begin{equation}
  \mathcal{F}
 = k_{B} T M \left[ - \frac{1}{\xi} + \frac{1}{\xi} e^{-\xi}
  - \frac{\epsilon}{k_{B} T} + \ln \frac{\xi}{\mathcal{Q}}\right]
\end{equation}
where $\xi$ is related to $M$, $V$ and $\epsilon$ via eq
\eqref{number_of_chains_chain_cluster_approximation}.

The number of slip-links $\tilde{Z}$ and the pressure $P$ are then
calculated to be
\begin{equation}
 \label{number_of_entanglements_chain_cluster_approximation}
 \tilde{Z}
  = 2 \left[ 1 - \frac{1 - e^{-\xi}}{\xi} \right]
\end{equation}
\begin{equation}
 \label{pressure_chain_cluster_approximation}
  P
  = \frac{k_{B} T M}{V} \frac{2 - \tilde{Z}}{2}
\end{equation}
From eq \eqref{number_of_entanglements_chain_cluster_approximation} we
find that $\tilde{Z}$ is a monotonically increasing function of $\xi$,
and at the limit of $\xi \to \infty$ (this corresponds to the strong
slip-linking limit), $\tilde{Z}$
approaches to $2$.
We also find that $P$ is a monotonically decreasing function of
$\tilde{Z}$, and thus as the slip-linking effect becomes strong, the
pressure decreases.
From eqs \eqref{pressure_ideal} and \eqref{pressure_chain_cluster_approximation},
we have eq \eqref{pressure_chain_cluster_approximation_final}. As
discussed in the main text, the pressure approaches to zero at the
strong slip-linking limit.

%------------------------------------------------------------------------------

%\clearpage

%%%%%%%%%%%%%%%%%%%%%%%%%%%%%%%%%%%%%%%%%%%%%%%%%%%%%%%%%%%%%%%%%%%%%%%%%%%%%%%%%
% BIBLIOGRAPHY

%%%%%%%%%%%%%%%%%%%%%%%%%%%%%%%%%%%%%%%%%%%%%%%%%%%%%%%%%%%%%%%%%%%%%%%%%%%%%%%%%

\clearpage
%%%%%%%%%%%%%%%%%%%%%%%%%%%%%%%%%%%%%%%%%%%%%%%%%%%%%%%%%%%%%%%%%%%%%%%%%%%%%%%%%
% FIGURE CAPTIONS

%%%%% FIGURE 1 ---- weakly_sliplinked_chains_image.eps
%%%%%%%%%%%%%%%%%%%%%%%%%%%%

%%%%% FIGURE 2 ---- chain_topologies_entanglement_image.eps
%%%%%%%%%%%%%%%%%%%%%%%%%%%%

%%%%% FIGURE 3 ---- chain_topologies_self_entanglement_image.eps
%%%%%%%%%%%%%%%%%%%%%%%%%%%%

%%%%%%%%%%%%%%%%%%%%%%%%%%%%%%%%%%%%%%%%%%%%%%%%%%%%%%%%%%%%%%%%%%%%%%%%%%%%%%%%%
% FIGURE FILES

\clearpage

%%\vspace*{0.1in}   %%% FIGURE 1
%\begin{center}
%\includegraphics[width=0.9\columnwidth,keepaspectratio=true]{weakly_sliplinked_chains_image.eps}
%\end{center}
%\vspace{0.25in}
%\hspace*{3in}
%{\Large
%\begin{minipage}[t]{3in}
%\baselineskip = .5\baselineskip
%Figure 1 \\
%Takashi Uneyama and Kazushi Horio \\
%J.\ Polym.\ Sci.\ B
%\end{minipage}
%}

%%\vspace*{0.1in}   %%% FIGURE 2
%\begin{center}
%\includegraphics[width=0.9\columnwidth,keepaspectratio=true]{chain_topologies_entanglement.eps}
%\end{center}
%\vspace{0.25in}
%\hspace*{3in}
%{\Large
%\begin{minipage}[t]{3in}
%\baselineskip = .5\baselineskip
%Figure 2 \\
%Takashi Uneyama and Kazushi Horio \\
%J.\ Polym.\ Sci.\ B
%\end{minipage}
%}

%%\vspace*{0.1in}   %%% FIGURE 3
%\begin{center}
%\includegraphics[width=0.9\columnwidth,keepaspectratio=true]{chain_topologies_self_entanglement.eps}
%\end{center}
%\vspace{0.25in}
%\hspace*{3in}
%{\Large
%\begin{minipage}[t]{3in}
%\baselineskip = .5\baselineskip
%Figure 3 \\
%Takashi Uneyama and Kazushi Horio \\
%J.\ Polym.\ Sci.\ B
%\end{minipage}
%}

%------------------------------------------------------------------------------
\end{document}